\documentclass[preprint,tightenlines,showpacs,amsmath,amssymb]{revtex4}

\usepackage{epsfig,epsf,graphics,psfrag}
\usepackage{times}
\usepackage{float}
\usepackage{color}
\usepackage{amsfonts,amssymb,stmaryrd,latexsym,amsmath}
\usepackage{multirow}
\usepackage{array} % for extrarowheigh
\usepackage{enumerate}

\begin{document}

\bibliographystyle{unsrt}

\title{Searching for charmoniumlike states with hidden $s\bar{s}$}

\author{Xiao-Hai Liu$^1$\footnote{liuxh@th.phys.titech.ac.jp} and Makoto Oka$^{1, 2}$\footnote{oka@th.phys.titech.ac.jp}}

\affiliation{ $^1$Department of Physics, H-27, Tokyo Institute of Technology, Meguro, Tokyo 152-8551, Japan}

\affiliation{$^2$ Advanced Science Research Center, JAEA, Tokai, Ibaraki 319-1195, Japan}

\date{\today}

\begin{abstract}
We investigate the processes $e^+e^-$$\to$$\gamma J/\psi\phi$, $\gamma J/\psi\omega$ and $\pi^0 J/\psi\eta$ to search for the charmnium-like states with hidden $s\bar{s}$, such as $Y(4140)$, $Y(4274)$, $X(4350)$ and $X(3915)$. These processes will receive contributions from the charmed-strange meson rescatterings. When the center-of-mass energies of the $e^+e^-$ scatterings are taken around the $D_{s0}(2317)D_s^{*}$, $D_{s1}(2460)D_s$ or $D_{s1}(2460)D_s^{*}$ threshold, the anomalous triangle singularities can be present in the rescattering amplitudes, which implies a non-resonance explanation about the resonance-like structures. The positions of the anomalous triangle singularities are sensitive to the kinematics, which offers us a criterion to distinguish the kinematic singularities from genuine particles.

%13.25.Gv  Decays of J/¦×, ¦´, and other quarkonia
%14.40.Rt	Exotic mesons
%12.39.Mk	Glueball and nonstandard multi-quark/gluon states

\end{abstract}

\maketitle

\section{Introduction}
With the number of charmoniumlike and bottomoniumlike $XYZ$ states observed in experiments increasing, the study on the exotic hadron spectroscopy is experiencing a renaissance in recent years. In the aspect of theory, most of these $XYZ$ states do not fit into the conventional quark model very well, which has been proved to be very successful in describing the heavy quarkonia below the open flavor thresholds. Various theoretical interpretations are then proposed to try to understand the underlying structures of these $XYZ$ states, such as hadronic molecule, tetraquark, hybrid, hadro-quarkonium, rescattering effect and so on. We refer to Refs.~\cite{Olsen:2014qna,Esposito:2014rxa,YI:2013vba} for both experimental and theoretical reviews about the $XYZ$ states.

In this work, we will focus on the exotic candidates which may contain the $s\bar{s}$ components, i.e. $Y(4140)$, $Y(4274)$, $X(4350)$ and $X(3915)$.
$Y(4140)$ and $Y(4274)$ were firstly observed by the CDF Collaboration in the $J/\psi\phi$ invariant mass distribution from $B\to K J/\psi\phi$ decays \cite{Aaltonen:2009tz,Aaltonen:2011at}. The presence of $Y(4140)$ in $B$ decays was later confirmed by the CMS and D0 Collaborations~\cite{Chatrchyan:2013dma,Abazov:2013xda,Abazov:2015sxa}. $X(4350)$ was observed by the Belle Collaboration from the two photon process $\gamma\gamma$$\to$$J/\psi\phi$ \cite{Shen:2009vs}. $Y(4140)$ and $Y(4274)$ were also expected to be produced in the two photon fusion reaction, but neither of them was observed \cite{Shen:2009vs}.
These resonance-like structures observed in the $J/\psi\phi$ mass spectrum are very intriguing, since they may contain both a $c\bar{c}$ pair and and an $s\bar{s}$ pair. Although their masses are well beyond the open charm thresholds, their widths are very narrow, for instance, $\Gamma_{Y(4140)}$$=$$15.3^{+10.7}_{-6.6}$ MeV, $\Gamma_{Y(4274)}$$=$$32.3^{+23.2}_{-17.1}$ MeV,
and $\Gamma_{Y(4350)}$$=$$13^{+18}_{-10}$ MeV. The above properties imply that these three states may be exotic. Taking into account their masses and decay modes, some people think $Y(4140)$, $Y(4274)$ and $X(4350)$ are probably the hadronic bound states of $D_s^{*+}D_s^{*-}$, $D_{s0}^{+}D_s^{-}$ and $D_{s0}^{+}D_s^{*-}$ respectively \cite{Liu:2009ei,Shen:2010ky,Liu:2010hf,He:2011ed,Finazzo:2011he,Wang:2011uk,Wang:2009wk,Wang:2014gwa,Albuquerque:2009ak,Ma:2014ofa,Ma:2014zva}. The tetraquark state $c\bar{c}s\bar{s}$ is also a possible explanation about them \cite{Stancu:2009ka,Wang:2015pea}.
 However, because of the low statistics, the masses and widths of these states still have larger uncertainties, even their existence are not well confirmed by different experiments \cite{YI:2013vba,Aaij:2012pz}.
Concerning $X(3915)$, it is observed in the $J/\psi\omega$ invariant mass distribution from both the $B$ decays $B\to K J/\psi\omega$ and the two photon fusion reaction $\gamma\gamma$$\to$$J/\psi\omega$. Although $X(3915)$ is currently taken as the conventional charmonium $\chi_{c0}(2P)$ by PDG \cite{Agashe:2014kda}, there are still some serious problems about this assignment. For instance, $X(3915)$ has not been observed in the $D\bar{D}$ invariant mass distribution, but the $D\bar{D}$ channel is expected to be the most important decay mode of $\chi_{c0}(2P)$. Furthermore, if $X(3915)$ is $\chi_{c0}(2P)$, the mass splitting between the well established $\chi_{c2}(2P)$, of which the mass is about 3927 MeV, and $\chi_{c0}(2P)$ is too small, which is in conflict with the theoretical predictions \cite{Olsen:2014qna,Godfrey:1985xj,Barnes:2005pb,Zhou:2015uva}. The width of $X(3915)$ is also very narrow, which is about 20 MeV. We notice that the mass threshold of $D_s^+D_s^-$ is about 3937 MeV, which is less than $J/\psi\phi$ threshold but close to $X(3915)$. Since there is a small $s\bar{s}$ component in the physical $\omega$ meson, we may wonder whether there are some connections between $D_s^+D_s^-$ system and $X(3915)$. In Ref.~\cite{Li:2015iga}, the authors suggest that $X(3915)$ may be the bound state of $D_s^+D_s^-$.

Before we claim these $XYZ$ states are genuine particles, such as molecule, tetraquark or hybrid, it is necessary to study some other possibilities. Some non-resonance explanations are also proposed to connect the "resonance-like" peaks, i.e. $XYZ$ states, with the kinematic singularities induced by the rescattering effects ~\cite{Chen:2011pv,Chen:2011zv,Bugg:2011jr,Chen:2011xk,Wang:2011yh,Wu:2011yx,Wang:2013cya,Liu:2013vfa,Liu:2014spa,Szczepaniak:2015eza,Liu:2015fea}. It is shown that sometimes it is not necessary to introduce a genuine resonance to describe a resonance-like structure, because some kinematic singularities of the rescattering amplitudes will behave themselves as bumps in the invariant mass distributions. The similar mechanism actually has been studied many years ago, such as the Peierls mechanism proposed in 1960s \cite{Peierls:1961zz,Goebel:1964zz,hwa:1963aa,landshoff:1962aa}. 
In this paper, we are going to investigate the correlations between the kinematic singularities and some exotic charmonium-like states with hidden $s\bar{s}$.

\section{Kinematic Singularity and Its Observable Phenomena}

\subsection{Radiative transitions}

\begin{figure}[htb]
	\centering
	\includegraphics[width=0.8\hsize]{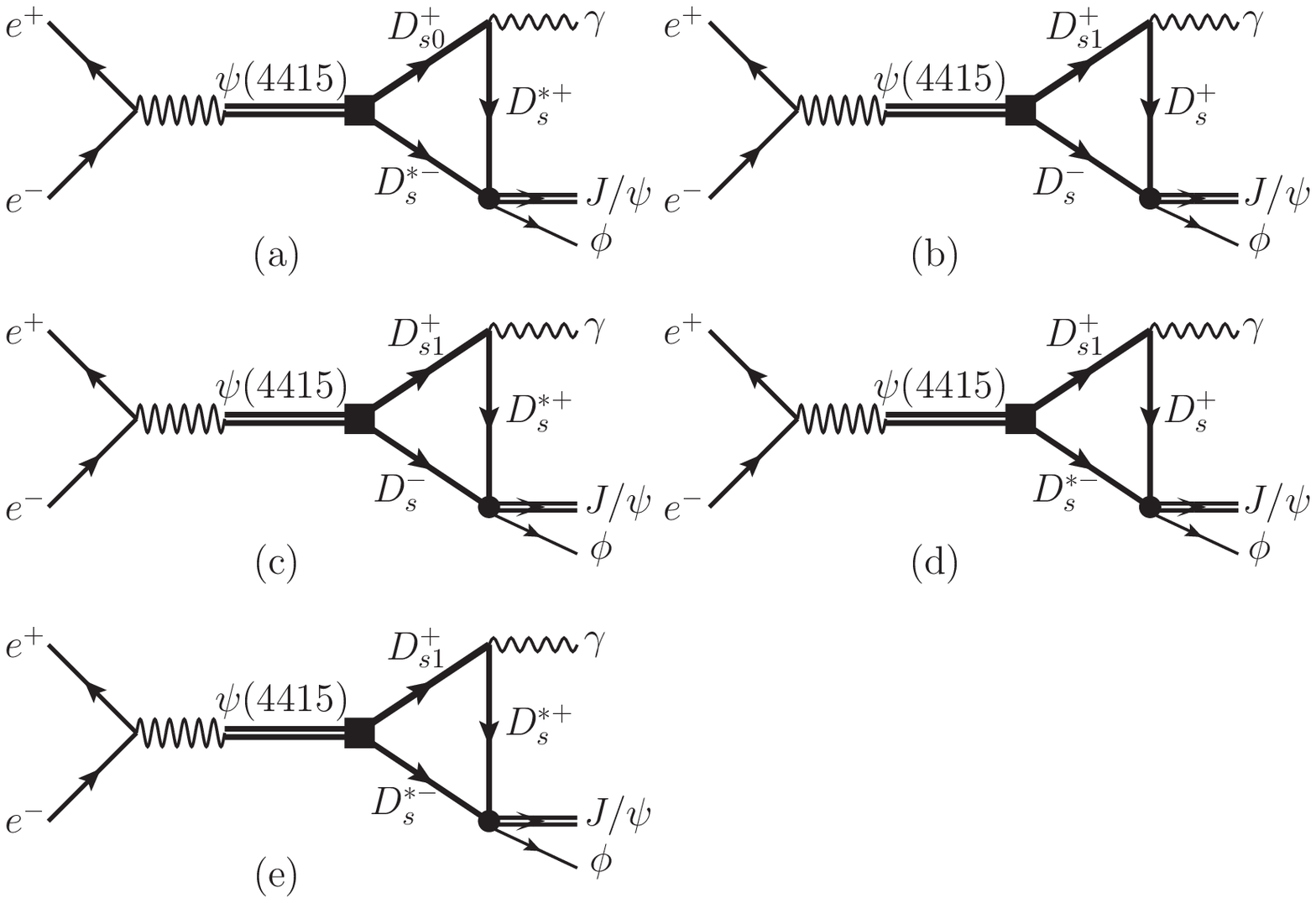}
	\caption{$e^+e^-$ scattering into $\gamma J/\psi\phi$ via $\psi(4415)$ and the charmed-strange meson rescattering loops.}\label{radiativedecay}
\end{figure}

\begin{figure}[t]
	\centering
	\includegraphics[width=0.38\hsize]{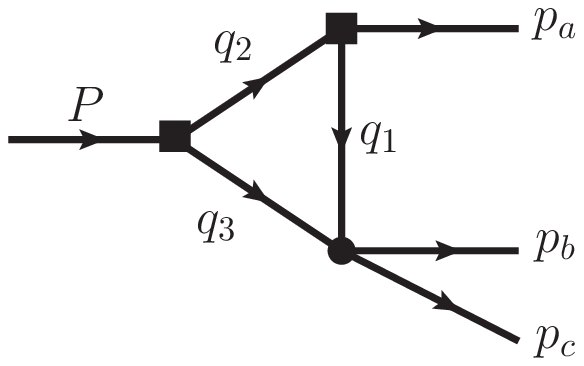}
	\caption{Triangle diagram under discussion. The internal mass which corresponds to the internal momentum $q_i$ is $m_i$ ($i$$=$1, 2, 3). For the external momenta, we define $P^2=s_1$, $(p_b+p_c)^2=s_2$ and $p_a^2=s_3$. We will use the same momentum and mass conventions in Figs.~\ref{radiativedecay} and \ref{jpsietapi}.}\label{triangle}
\end{figure}

Since the charmonium-like states with hidden $s\bar{s}$ may decay into $J/\psi\phi$, besides the $B$ decays and the two photon fusion reactions, we also hope to search for these states in the process $e^+e^-$$\to$$\gamma J/\psi\phi$, taking into account the high statics of the modern experimental facilities, such as BESIII and Belle. The process $e^+e^-$$\to$$\gamma J/\psi\phi$ will receive contributions from the rescattering diagrams as displayed in Fig.~\ref{radiativedecay}. There are several reasons why we expect the rescatterings induced by these charmed-strange meson loops will be important. Firstly, $\psi(4415)$ is widely accepted as the $S$-wave charmonium $\psi(4S)$, and it can couple to $D_{s0}(2317)D_s^{*}$ ($D_{s1}(2460)D_s$, $D_{s1}(2460)D_s^{*}$) in relative $S$-wave. This $S$-wave coupling will respect the heavy quark spin symmetry (HQSS). The quark model calculation also indicates that this coupling will be strong. Secondly, since $M_{\psi(4415)}$ is very close to the mass thresholds of $D_{s0}(2317)D_s^{*}$ and $D_{s1}(2460)D_s$, if we collect the data samples at the center-of-mass (CM) energies near $\psi(4415)$, an intriguing kinematic singularity, i.e., the anomalous triangle singularity (ATS), may emerge in the rescattering amplitude. Another important reason is all of the internal charmed-strange mesons appeared in the rescattering diagrams are very narrow \cite{Agashe:2014kda}, which implies that the effect of the occurrence of the ATS will be obvious \cite{Liu:2015taa}.

The ATS corresponds to a pinch singularity of the loop integral. In Ref.~\cite{Liu:2015taa}, we have discussed the kinematic conditions under which the ATS can be present. Taking into account the Feynman diagram displayed in Fig.~\ref{triangle}, according to the single dispersion representation of the triangle diagram, the locations of the ATS for $s_1$ and $s_2$ can be determined as 
\begin{eqnarray}
s_1^{-}&=&(m_2+m_3)^2+\frac{1}{2m_1^2} {\LARGE[}(m_1^2+m_2^2-s_3)(s_2-m_1^2-m_3^2)-4m_1^2 m_2 m_3 \nonumber \\  &-& \lambda^{1/2}(s_2,m_1^2,m_3^2)\lambda^{1/2}(s_3,m_1^2,m_2^2){\LARGE ]},
\end{eqnarray}
and
\begin{eqnarray}
s_2^{-}&=&(m_1+m_3)^2+\frac{1}{2m_2^2} {\LARGE[}(m_1^2+m_2^2-s_3)(s_1-m_2^2-m_3^2)-4m_2^2 m_1 m_3 \nonumber \\  &-& \lambda^{1/2}(s_1,m_2^2,m_3^2)\lambda^{1/2}(s_3,m_1^2,m_2^2){\LARGE ]},
\end{eqnarray}  
respectively, where $\lambda(x,y,z)\equiv (x-y-z)^2-4yz$. $s_1^-$ and $s_2^-$ are the so-called anomalous thresholds. For convenient, we define the normal threshold $s_{1N}$ ($s_{2N}$) and the critical value $s_{1C}$ ($s_{2C}$) for $s_1$ ($s_2$) as follows,
\begin{eqnarray}\label{s1Ns1C}
&& s_{1N}=(m_2+m_3)^2,\ s_{1C}=(m_2+m_3)^2 +\frac{m_3}{m_1}[(m_2-m_1)^2-s_3], \\
&& s_{2N}=(m_1+m_3)^2,\ s_{2C}=(m_1+m_3)^2 +\frac{m_3}{m_2}[(m_2-m_1)^2-s_3].
\end{eqnarray}
If we fix $s_3$ and the internal masses $m_{1,2,3}$, when $s_1$ increases from $s_{1N}$ to $s_{1C}$, the anomalous threshold $s_2^-$ will move from $s_{2C}$ to $s_{2N}$. Likewise, when $s_2$ increases from $s_{2N}$ to $s_{2C}$, $s_1^-$ will move from $s_{1C}$ to $s_{1N}$. This is the kinematic region where the ATS can be present.
The discrepancies between the normal and anomalous thresholds are defined as follows,
\begin{eqnarray}
\Delta_{s_1}&=&\sqrt{s_1^-} - \sqrt{s_{1N}}, \nonumber \\
\Delta_{s_2}&=&\sqrt{s_2^-} - \sqrt{s_{2N}}.
\end{eqnarray}
Apparently, when $s_2$$=$$s_{2N}$ ($s_1$$=$$s_{1N}$), we will obtain the maximum value of $\Delta_{s_1}$ ($\Delta_{s_2}$), i.e.,
\begin{eqnarray} \label{deltas1s2}
\Delta_{s_1}^{\mbox{max}}&=&\sqrt{s_{1C}} - \sqrt{s_{1N}}, \\
\Delta_{s_2}^{\mbox{max}}&=&\sqrt{s_{2C}} - \sqrt{s_{2N}}.
\end{eqnarray}
Larger $\Delta_{s_1}^{\mbox{max}}$ and $\Delta_{s_2}^{\mbox{max}}$ indicate larger kinematic regions where the ATS can emerge, which also implies that it will be easier to detect the ATS in experiments. Notice that as long as $s_3$ and the internal masses $m_{1,2,3}$ are fixed,  $\Delta_{s_1}^{\mbox{max}}$ and $\Delta_{s_2}^{\mbox{max}}$ are determined. The corresponding $\Delta_{s_1}^{\mbox{max}}$ and $\Delta_{s_2}^{\mbox{max}}$ of the diagrams in Fig.~\ref{radiativedecay} are listed in Table~\ref{tabdeltarad}. From Table~\ref{tabdeltarad}, we can see that although $\Delta_{s_1}^{\mbox{max}}$ and $\Delta_{s_2}^{\mbox{max}}$ are not very large, they are still sizable. This is because the phase spaces for $D_{s0}^{\pm}(2317)$$\to$$\gamma D_s^{*\pm}$ and $D_{s1}^{\pm}(2460)$$\to$$\gamma D_s^{(*)\pm}$ are relatively larger, as discussed in Ref.~\cite{Liu:2015taa}.

\begin{table}
	\caption{$\Delta_{s_1}^{\mbox{max}}$ and $\Delta_{s_2}^{\mbox{max}}$ for the corresponding triangle diagrams in Fig.~\ref{radiativedecay}.}
	\begin{center}
		\begin{tabular}{|c|c|c|c|c|c|}
			\hline [MeV] & Fig.~\ref{radiativedecay}(a) & Fig.~\ref{radiativedecay}(b) & Fig.~\ref{radiativedecay}(c) & Fig.~\ref{radiativedecay}(d) & Fig.~\ref{radiativedecay}(e) \\ 
			\hline $\Delta_{s_1}^{\mbox{max}}$ & 4.8 & 27 & 13  & 28 & 13 \\ 
			\hline $\Delta_{s_2}^{\mbox{max}}$ & 4.6  & 24  & 12 & 25 & 12 \\ 
			\hline 
		\end{tabular} 
	\end{center}\label{tabdeltarad}
\end{table}

The above kinematic analysis indicates that the ATS induced by the charmed-strange meson loops may emerge in a relatively larger kinematic region. To quantitatively estimate how important these rescattering amplitudes are, we will build our model in the framework of heavy hadron chiral
perturbation theory (HHChPT)~\cite{Casalbuoni:1992dx,Casalbuoni:1996pg,Colangelo:2003sa,Mehen:2011yh,Margaryan:2013tta,Guo:2013zbw,Mehen:2004uj}. In HHChPT, to encode the HQSS, the charmed meson doublets with light degrees of freedom
$J^P=1/2^-$ and $1/2^+$ are collected into the following superfields
\begin{eqnarray}
&& H_{1a}  = \frac{1+{\rlap{v}/}}{2}[\mathcal{D}_{a\mu}^*\gamma^\mu-\mathcal{D}_a\gamma_5] , \\
&& H_{2a}  = [\bar{\mathcal{D}}_{a\mu}^*\gamma^\mu+\bar{\mathcal{D}}_a\gamma_5]\frac{1-{\rlap{v}/}}{2} , \\
&& S_{1a} = \frac{1+{\rlap{v}/}}{2} \left[\mathcal{D}_{1a}^{\prime \mu}\gamma_\mu\gamma_5-\mathcal{D}_{0a}\right] , \\
&& S_{2a} =  \left[\bar{\mathcal{D}}_{0a}-\bar{\mathcal{D}}_{1a}^{\prime \mu}\gamma_\mu\gamma_5\right]\frac{1-{\rlap{v}/}}{2} \ ,\\
&&\bar{H}_{1a,2a}=\gamma^0 H_{1a,2a}^{\dag} \gamma^0,\ \bar{S}_{1a,2a}=\gamma^0 S_{1a,2a}^{\dag} \gamma^0 ,
\end{eqnarray}
where $H_{2a}$ ($S_{2a}$) is the charge conjugate field of $H_{1a}$ ($S_{1a}$), and $a$ is the light flavor index. We identify the physical states $D_{s0}^{\pm}(2317)$ and $D_{s1}^{\pm}(2460)$ as the doublet collected in the superfield $S_{1a,2a}$, which is widely accepted. The pertinent effective Lagrangian which respects the HQSS and chiral symmetry takes the form 
\begin{eqnarray}\label{lageff}
\mathcal{L}_{eff}
&=& g_S < J \bar{S}_{2a}  \bar{H}_{1a}  +  J \bar{H}_{2a}  \bar{S}_{1a} >  
+ C_P < J \bar{H}_{2b} \gamma_\mu\gamma_5 \bar{H}_{1a} \mathcal{A}^\mu_{ba} >  
+ C_V < J \bar{H}_{2b} \gamma_\mu \bar{H}_{1a} \mathcal{\rho}^\mu_{ba} > \nonumber \\
&+& ih < {\bar H}_{1a} S_{1b} \gamma_\mu \gamma_5 {\cal A}_{ba}^\mu > 
+ \frac{e\tilde{\beta}}{4} < {\bar H}_{1a} S_{1b} 
\sigma^{\mu\nu} F_{\mu\nu} Q_{ba}>\ ,
\end{eqnarray}
where $<\cdot\cdot\cdot>$ means the trace over Dirac matrices, $J$ indicates the $S$-wave charmonia
\begin{eqnarray}
J &=& \frac{1+{\rlap{v}/}}{2} [\psi(nS)^\mu \gamma_\mu-\eta_c(nS)\gamma_5]
\frac{1-{\rlap{v}/}}{2}\ ,
\end{eqnarray}
$\mathcal{A}^\mu$ is the chiral axial vector containing the
Goldstone bosons 
\begin{eqnarray}
\mathcal{A}_\mu &=& \frac{1}{2} \left( \xi^\dag\partial_\mu\xi  - \xi\partial_\mu\xi^\dag  \right)\ ,
\end{eqnarray}
with
\begin{eqnarray}
\xi = e^{i\mathcal{M}/f_\pi}\ ,\ 
\mathcal{M}= \left(
\begin{array}{ccc}
\frac{1}{\sqrt{2}}\pi^0+ \frac{1}{\sqrt{6}}\eta & \pi^+ & K^{+} \\
\pi^- & -\frac{1}{\sqrt{2}}\pi^0+ \frac{1}{\sqrt{6}}\eta  & K^{0} \\
K^{-} & \bar{K}^{0} & -\sqrt{\frac{2}{3}}\eta \\
\end{array}
\right),
\end{eqnarray}
$\rho^\mu$ is a $3\times 3$ matrix for the nonet vector mesons
\begin{eqnarray}\label{rhomatix}
{\rho}= \left(
\begin{array}{ccc}
\frac{1}{\sqrt{2}}\rho^0+ \frac{1}{\sqrt{2}}\omega & \rho^+ & K^{*+} \\
\rho^- & -\frac{1}{\sqrt{2}}\rho^0+ \frac{1}{\sqrt{2}}\omega  & K^{*0} \\
K^{*-} & \bar{K}^{*0} & \phi \\
\end{array}
\right),
\end{eqnarray}
$F_{\mu\nu}$ is the electromagnetic field tensor
\begin{eqnarray}
F_{\mu\nu}=\partial_\mu A_\nu-\partial_\nu A_\mu,
\end{eqnarray} 
and
\begin{eqnarray}
Q=\mbox{diag}(2/3,\ -1/3,\ -1/3).
\end{eqnarray}

The coupling constants $h$ and $\tilde{\beta}$ in Eq.(\ref{lageff}), which are in relevant with the strong and radiative decay rates of the charmed mesons respectively, can be extracted according to the experimental data. We will take the averaged values of $h$ and $\tilde{\beta}$  estimated in Ref.~\cite{Mehen:2004uj} in our following numerical calculations, which are
$h^2$=0.44 and $|\tilde{\beta}|$=0.42 GeV$^{-1}$ respectively. For the coupling constant $g_S$, by matching the decay amplitude of $\psi(4S)$$\to$$D_{s0}^+D_{s}^{*-}$ calculated according to Eq.~(\ref{lageff}) with that calculated in the quark pair creation model~\cite{Barnes:2005pb}, we obtain $g_S$$\approx$1.51 $\mbox{GeV}^{-1/2}$ . Similarly, by matching the scattering amplitudes of $D_s^{*+}D_s^{-}\to J/\psi\eta$ and $D_s^{*+}D_s^{*-}\to J/\psi\phi$ calculated according to Eq.~(\ref{lageff}) with those calculated in the quark-interchange model, we obtain $C_P$$\approx$1.73 $\mbox{GeV}^{-3/2}$ and $C_V$$\approx$46 $\mbox{GeV}^{-3/2}$ respectively. We give a brief introduction about the quark-interchange model in Appendix A. Of course the estimation of the coupling constants using quark model will be model-dependent, and may have relatively larger uncertainties, but we expect that the order of magnitude of this estimation is still reasonable to some extent. Notice that in HHChPT, every heavy filed $H$ will contain a factor $\sqrt{M_H}$ for normalization.

According to the effective Lagrangian in Eq.~\ref{lageff}, the transition amplitude of $\psi(4S)\to\gamma J/\psi\phi$ corresponding to the rescattering diagram Fig.~\ref{radiativedecay}(a) reads
\begin{eqnarray}\label{eqampA}
T^{A}_{\psi(4S)\to\gamma J/\psi\phi}&=& \frac{2}{3} g_S e \tilde{\beta} C_V \int \frac{d^4 q_1}{(2\pi)^4} \frac{1}{(q_1^2-M_{D_s^{*+}}^2)(q_2^2-M_{D_{s0}^{+}}^2)(q_3^2-M_{D_s^{*-}}^2)} \nonumber \\
&\times& \big( \epsilon_{\psi(4S)} \cdot  \epsilon_{J/\psi}^* \ v \cdot  \epsilon_{\gamma}^* \ p_a \cdot  \epsilon_{\phi}^* +
\epsilon_{\psi(4S)} \cdot  \epsilon_{\phi}^* \ v \cdot  \epsilon_{\gamma}^* \ p_a \cdot  \epsilon_{J/\psi}^* \nonumber \\
&+& \epsilon_{\psi(4S)} \cdot  \epsilon_{\gamma}^* \ \epsilon_{J/\psi}^* \cdot  \epsilon_{\phi}^* \ v \cdot  p_a -
\epsilon_{\psi(4S)} \cdot  \epsilon_{J/\psi}^* \ \epsilon_{\gamma}^* \cdot  \epsilon_{\phi}^* \ v \cdot  p_a
 \nonumber \\
&-& \epsilon_{\psi(4S)} \cdot  \epsilon_{\phi}^* \ \epsilon_{\gamma}^* \cdot  \epsilon_{J/\psi}^* \ v \cdot  p_a -
p_a \cdot  \epsilon_{\psi(4S)} \ v \cdot  \epsilon_{\gamma}^* \ \epsilon_{J/\psi}^* \cdot  \epsilon_{\phi}^* \big),
\end{eqnarray}
where $\epsilon_{\gamma}$, $\epsilon_{J/\psi}$, $\epsilon_{\phi}$, $\epsilon_{\psi(4S)}$ are the polarization vectors of the corresponding particles, and the velocity $v$ can be taken as $(1,0,0,0)$ in the static limit. The other transition amplitudes share the similar formula with Eq.~(\ref{eqampA}), which are omitted for brevity. We will introduce a Breit-Wigner type propagator of $\psi(4415)$ when calculating the scattering amplitude of $e^+e^-$$\to$$\gamma J/\psi\phi$ via $\psi(4415)$ and the charmed-strange meson loops. The propagator takes the form
\begin{equation}
BW[\psi(4415)]=(s_1-M_{\psi(4415)}^2+i M_{\psi(4415)}\Gamma_{\psi(4415)})^{-1}.
\end{equation}   
The coupling between the virtual photon and $\psi(4415)$ will be determined by means of the vector meson dominance model \cite{Bauer:1975bw,Li:2008xm,Zhang:2008ab}.

\begin{figure}[tb]
	\centering
	\includegraphics[width=0.6\hsize]{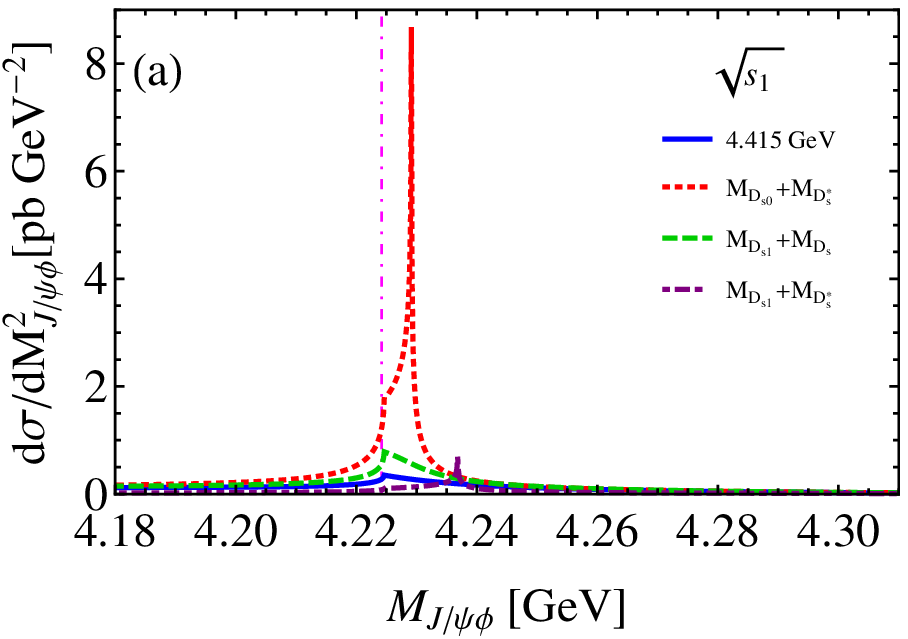}\\
	\includegraphics[width=0.6\hsize]{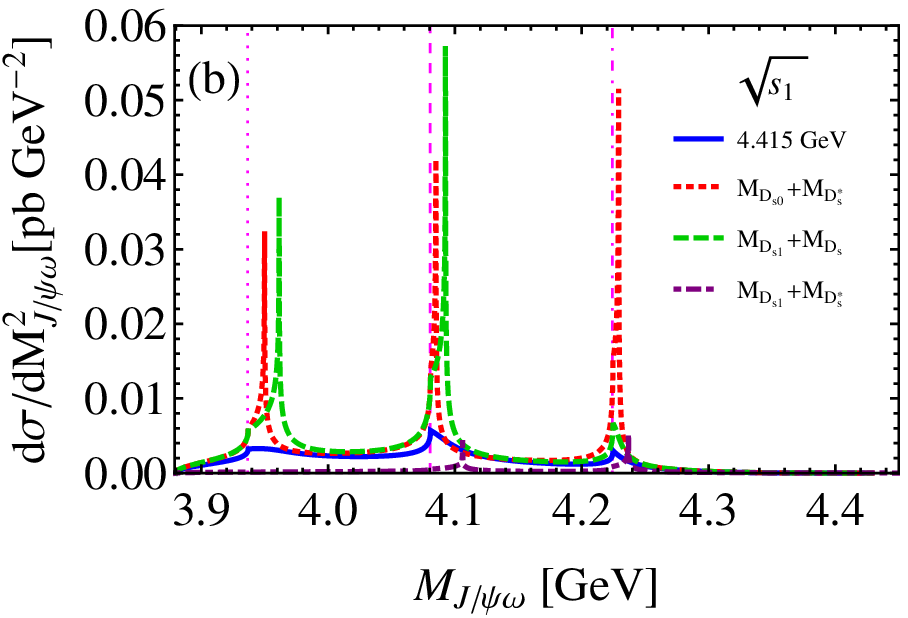}\\
	\includegraphics[width=0.6\hsize]{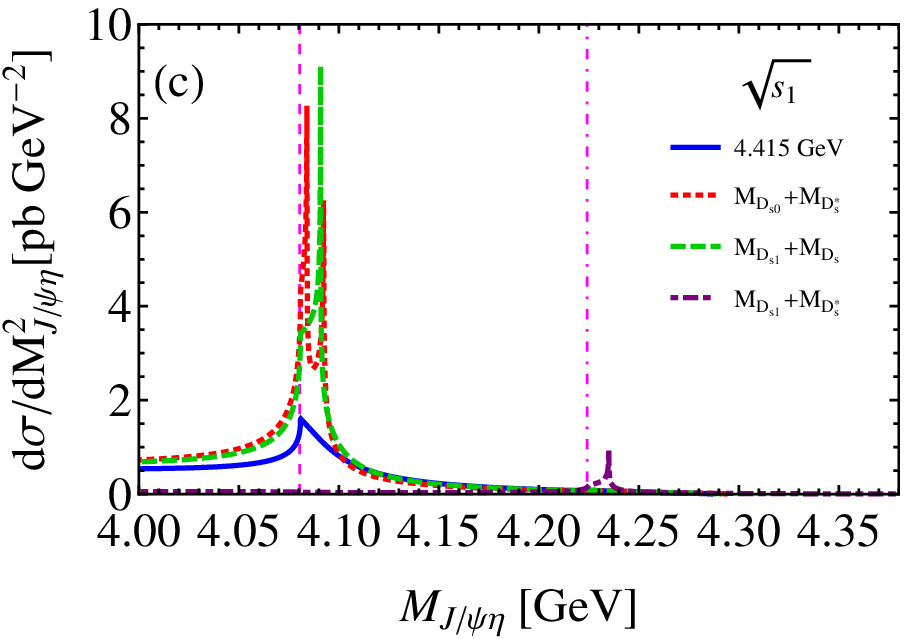}
	\caption{Invariant mass distributions of (a) $J/\psi\phi$, (b) $J/\psi\omega$, and (c) $J/\psi\eta$ at four CM energy points. The vertical dotted, dashed, dot-dashed grid lines indicate the $D_s^{+}D_s^{-}$, $D_s^{*+}D_s^{-}$, and $D_s^{*+}D_s^{*-}$ thresholds, respectively.}\label{invariantmass}
\end{figure}

The numerical results for the invariant mass distribution of $J/\psi\phi$ in the process $e^+e^-$$\to$$\gamma J/\psi\phi$ via the charmed-strange meson loops are displayed in Fig.~\ref{invariantmass}(a). We calculate the differential cross sections at several CM energies, i.e. 4.415 GeV and three thresholds. 4.415 GeV is actually out of the kinematic region where the ATS can be present, therefore when $\sqrt{s_1}$=4.415 GeV, there is only a small cusp appeared in the normal $D_s^{*+}D_s^{*-}$ threshold. Since the $J/\psi\phi$ threshold is only below the $D_s^{*+}D_s^{*-}$ threshold, but above the $D_s^{+}D_s^{-}$ and $D_s^{*+}D_s^{-}$  thresholds, according to Table~\ref{tabdeltarad}, among the five rescattering diagrams of Fig.~\ref{radiativedecay}, only in the rescattering amplitudes corresponding to Figs.~\ref{radiativedecay}(a) and (e), the ATS can appear in the physical kinematic region. When the CM energy $\sqrt{s_1}$ is taken at the $D_{s1}^{+}D_s^{-}$ threshold, since the ATS can not be present in the rescattering amplitudes corresponding to Figs.~\ref{radiativedecay}(b) and (c), there is only a small cusp stay at $D_s^{*+}D_s^{*-}$ threshold (dashed line in Fig.~\ref{invariantmass}(a)).  When $\sqrt{s_1}$=$M_{D_{s0}}$+$M_{D_{s}^*}$, the ATS will be present in the rescattering amplitude corresponding to Fig.~\ref{radiativedecay}(a), which lies about 4.6 MeV above the $D_s^{*+}D_s^{*-}$ threshold, as illustrated in Fig.~\ref{invariantmass}(a) (dotted line). When $\sqrt{s_1}$=$M_{D_{s1}}$+$M_{D_{s}^*}$, the ATS will be present in the rescattering amplitudes corresponding to Fig.~\ref{radiativedecay}(e), which lies about 12 MeV above $D_s^{*+}D_s^{*-}$ threshold (dot-dashed line in Fig.~\ref{invariantmass}(a)). However, the CM energy $\sqrt{s_1}$=$M_{D_{s1}}$+$M_{D_{s}^*}$ is far away from the peak position of the resonance $\psi(4415)$, in which case the contribution of the diagram Fig.\ref{radiativedecay}(e) will be suppressed to some extent.

Notice that the resonance-like peaks appeared in Fig.~\ref{invariantmass} are not induced by any genuine resonances, and the peak positions and shapes are very sensitive to the kinematics. As we point out in Ref.~\cite{Liu:2015taa}, the difference between the genuine particles and the kinematic singularities is that the resonance-like peaks induced by the kinematic singularities will depend on the kinematic configurations, which means that the peak positons of the resonance-like structures will depend on the production modes. 

The estimated cross section of the process is of the order of magnitude of 1 pico barn. With the huge statics of the modern experimental facilities, the effects induced by the ATS may be detectable at BESIII, Belle or the forthcoming Belle II.

As mentioned above, the higher $J/\psi\phi$ threshold leads to that only the ATS which is in relevant with the $D_s^{*+}D_s^{*-}$ threshold can emerge in the process $e^+e^-\to \gamma J/\psi\phi$. On the other hand, the $J/\psi\omega$ threshold is even below the $D_s^{+}D_s^{-}$ threshold, and $D_s^{(*)+}D_s^{(*)-}$ can also scatter into $J/\psi\omega$,  which imply that there will be three ATSs in relevant with three thresholds can be present in the rescattering amplitudes of $e^+e^-\to \gamma J/\psi\omega$. However, because there is only a small $s\bar{s}$ component in $\omega$, the scattering amplitudes of $D_s^{(*)+}D_s^{(*)-}$$\to$$J/\psi\omega$ will be suppressed. 

We will estimate the amplitudes of $e^+e^-\to \gamma J/\psi\omega$ via charmed-strange meson loops by taking into account the $\phi$-$\omega$ mixing.   
When we introduce the vector nonet matrix $\rho_\mu$ in Eqs.~(\ref{lageff}) and (\ref{rhomatix}), we have assumed an ideal mixing between the flavor singlet and octet. The physical states $\phi$ and $\omega$ are actually not pure $s\bar{s}$
and $(u\bar{u}+d\bar{d})/\sqrt{2}$, respectively. We rewrite their wave functions as follows: 
\begin{eqnarray}
\phi&=&\mbox{sin}\theta_{\phi\omega}\ (u\bar{u}+d\bar{d})/\sqrt{2}-\mbox{cos}\theta_{\phi\omega}\ s\bar{s},\\
\omega&=&\mbox{cos}\theta_{\phi\omega}\ (u\bar{u}+d\bar{d})/\sqrt{2}+\mbox{sin}\theta_{\phi\omega}\ s\bar{s},
\end{eqnarray}
where the mixing angle $\theta_{\phi\omega}$ is approximately equal to 0.065, by means of the quadratic Gell-Mann-Okubo mass formula~\cite{Okubo:1961jc,Okubo:1962zzc,Kucukarslan:2006wk,Cheng:2011fk}. The numerical results of $J/\psi\omega$ invariant mass distributions are displayed in Fig.~\ref{invariantmass}(b). Being different from Fig.~\ref{invariantmass}(a), for some CM energies, there are three peaks staying in the vicinities of $D_s^{+}D_s^{-}$, $D_s^{*+}D_s^{-}$ and $D_s^{*+}D_s^{*-}$ thresholds respectively. Furthermore, according to Table~\ref{tabdeltarad}, it seems that in a relatively larger kinematic region these resonance-like peaks can be observed. However, compared with the process 
$e^+e^-\to \gamma J/\psi\phi$, the cross section of $e^+e^-\to \gamma J/\psi\omega$ via charmed-strange meson loops is nearly suppressed by two orders of magnitude, which will make the observation of the peaks induced by the ATS become difficult. The process $e^+e^-\to \gamma J/\psi\omega$ will also receive contributions from other rescattering diagrams, such as the $D_0\bar{D}^*D$ and $D_1^\prime\bar{D}D^*$ loops. But because $D_0$ and $D_1^\prime$ are too broad, the rescattering amplitudes will be highly suppressed and can only be taken as the backgrounds, as discussed in Refs.~\cite{Liu:2013vfa,Liu:2014spa}. 

The BESIII Collaboration has ever searched for the charmonium-like state $Y(4140)$ in the process $e^+e^-\to \gamma J/\psi\phi$, but no significant signal is observed~\cite{Ablikim:2014atq}. This result can be understood in our scenario. Firstly, if $Y(4140)$ is not a genuine particle but the kinematic threshold effect, it is not strange that people observe it in $B$ decays rather than in $e^+e^-$ scatterings,  because of the different kinematic configurations in these two reactions.  Secondly, the BESIII Collaboration used the data samples collected at the CM energies 4.23, 4.26 and 4.36 GeV, but unfortunately none of these CM energies falls into the kinematic regions where the ATS can be present according to Table~\ref{tabdeltarad}. When the CM energies are taken in the range 4.430$\sim$4.435 GeV or 4.572$\sim$4.585 GeV according to Table~\ref{radiativedecay}, one may probably observe some resonance-like peaks in $J/\psi\phi$ invariant mass distributions, which are induced by the ATS. However, in our numerical results Fig~\ref{invariantmass}(a), there are only peaks staying close to the $D_s^{*+}D_s^{*-}$ threshold, which are somewhat far away from the peak position of $Y(4140)$. Since the kinematics and rescattering processes in $B$ decays will be another story, here we can only point out the possibility but can not verify that the production of $Y(4140)$ is induced by the kinematic threshold effect.

\subsection{Isospin-symmetry breaking process}

\begin{figure}[htb]
	\centering
	\includegraphics[width=0.8\hsize]{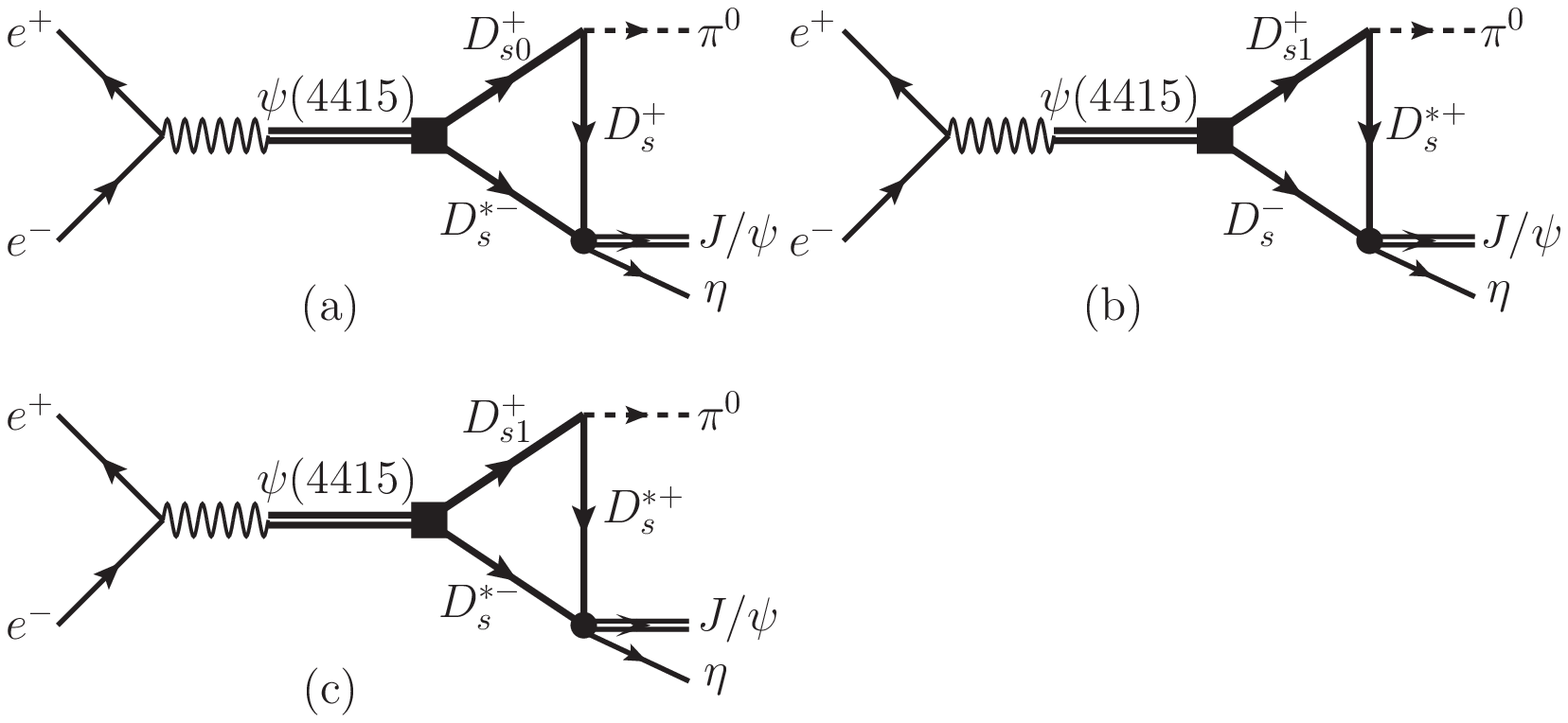}
	\caption{$e^+e^-$ scattering into $\pi^0 J/\psi\eta$ via $\psi(4415)$ and the charmed-strange meson rescattering loops.}\label{jpsietapi}
\end{figure}

\begin{table}
	\caption{$\Delta_{s_1}^{\mbox{max}}$ and $\Delta_{s_2}^{\mbox{max}}$ for the corresponding triangle diagrams in Fig.~\ref{jpsietapi}.}
	\begin{center}
		\begin{tabular}{|c|c|c|c|c|c|}
			\hline [MeV] & Fig.~\ref{jpsietapi}(a) & Fig.~\ref{jpsietapi}(b) & Fig.~\ref{jpsietapi}(c) \\ 
			\hline $\Delta_{s_1}^{\mbox{max}}$ & 13 & 11 & 11 \\ 
			\hline $\Delta_{s_2}^{\mbox{max}}$ & 12  & 10  & 10 \\ 
			\hline 
		\end{tabular} 
	\end{center}\label{tabjpsieta}
\end{table}

The $C$-parity of the $J/\psi\phi$ or $J/\psi\omega$ combination must be positive, but for the $D_s^{(*)+}D_s^{(*)-}$ combination, the $C$-parity can either be positive or negative. In our scenario, we suppose that the resonance-like peaks observed in $J/\psi\phi$ ($J/\psi\omega$) can be related with the rescattering loops which contain the vertices of $D_s^{(*)+}D_s^{(*)-}$ scattering into $J/\psi\phi$ ($J/\psi\omega$). Likewise, we can also expect the similar peaks in other final states with the negative $C$-parity. For instance, since $D_s^{(*)+}D_s^{(*)-}$ can also scatter into $J/\psi\eta$, of which the $C$-parity is negative, we can then study the process $e^+e^-\to \pi^0 J/\psi\eta$ via the charmed-strange meson loops. In another word, we hope to search for the negative $C$-parity charmonium-like structures with hidden $s\bar{s}$ in $e^+e^-\to \pi^0 J/\psi\eta$. 

The diagrams for $e^+e^-\to \pi^0 J/\psi\eta$ via the charmed-strange meson loops are displayed in Fig.~\ref{jpsietapi}. Notice that in these diagrams, there are vertices for $D_{s0}^\pm$ coupling to $D_s^\pm\pi^0$ and $D_{s1}^\pm$ coupling to $D_s^{*\pm}\pi^0$. Although the isospin symmetry is not conserved in these couplings, the processes $D_{s0}^\pm$$\to$$D_s^\pm\pi^0$
and $D_{s1}^\pm$$\to$$D_s^{*\pm}\pi^0$ are acturally the most important decay modes for $D_{s0}^\pm$ and $D_{s1}^\pm$ respectively. This is because the isospin conserved $DK$ and $D^*K$ channels are not open for these two $P$-wave charmed strange mesons. We therefore expect the rescattering amplitudes corresponding to Fig.~\ref{jpsietapi} will be important for $e^+e^-\to \pi^0 J/\psi\eta$. To estimate the amplitudes, we will use the effective Lagrangian in Eq.~(\ref{lageff}). The decays $D_{s0}^\pm$$\to$$D_s^\pm\pi^0$
and $D_{s1}^\pm$$\to$$D_s^{*\pm}\pi^0$ can proceed via the $\eta$-$\pi^0$ mixing,  
and the mixing angle $\theta_{\eta\pi}$=$\sqrt{3}/2(m_d-m_u)/(2m_s-m_d-m_u)$$\simeq$0.01, which is a widely accepted value.

The kinematic regions where the ATS can be present are listed in Table~\ref{tabjpsieta}. Because the phase spaces of $D_{s0}^\pm$$\to$$D_s^\pm\pi^0$
and $D_{s1}^\pm$$\to$$D_s^{*\pm}\pi^0$ are larger, the kinematic regions for the occurrence of the ATS are also relatively larger. The numerical results of the $J/\psi\eta$ invariant mass distributions are displayed in Fig.~\ref{invariantmass}(c). Due to the rescattering diagrams in Fig.~\ref{jpsietapi}, when the CM energy $\sqrt{s_1}$ is smaller than the $D_{s1}D_s^*$ threshold, only the peaks close to the $D_s^{*+}D_s^{-}$ threshold can appear. There is an intriguing property for the lineshapes of the invariant mass distributions. When the CM energy $\sqrt{s_1}$ is taken to be the $D_{s0}D_s^*$ threshold (dotted line in Fig.~\ref{invariantmass}(c)), taking into account Table.~\ref{tabjpsieta}, the kinematic regions of the ATS corresponding to Figs.~\ref{jpsietapi}(a) and (b) will overlap, but the locations of the ATSs for $\sqrt{s_2}$ ($J/\psi\eta$ invariant mass) are different. Therefore when $\sqrt{s_1}$=$M_{D_{s0}}$$+$$M_{D_s^*}$, there will be two peaks appeared in the invariant mass distribution, and both of them stay close to the $D_s^{*+}D_s^{-}$ threshold. Notice that there will be no peaks staying close to the $D_s^{+}D_s^{-}$ threshold, because there is no vertex for $D_s^{+}D_s^{-}$$\to$$J/\psi\eta$ included in the diagrams of Fig.~\ref{jpsietapi}.

The cross sections of this isospin-symmetry breaking process are estimated at the order of magnitude of 1 pico barn. In Ref.~\cite{Ablikim:2015xfo}, the BESIII Collaboration reports some results about the cross sections $\sigma$($e^{+}e^{-}$$\to$$\pi^0J/\psi\eta$), of which the upper limits are also at the order of magnitude of 1 pico barn. However, for the CM energies where the data are collected in Ref.~\cite{Ablikim:2015xfo}, none of them falls into the kinematic regions where the ATS can be present according to Table~\ref{tabjpsieta}. To observe the resonance-like peaks induced by the ATS, maybe one should collect the data at other CM energies, of which the range is 4.428$\sim$4.443 GeV or 4.572$\sim$4.583 GeV.   

The process $e^{+}e^{-}$$\to$$\pi^0J/\psi\eta$ will also receive contributions from other charmed meson rescattering diagrams, such as the $D_1\bar{D}D^*$ loop, which has been estimated in Ref.~\cite{Wu:2013onz}. However, taking into account that $D_1$ is much broader than $D_{s1}$ and $D_{s0}$, and the scattering $e^+e^-$$\to$$D_1\bar{D}$ will be suppressed by the HQSS \cite{Liu:2014spa,Li:2013yka,Cleven:2013mka}, we suppose that the contribution for $e^{+}e^{-}$$\to$$\pi^0J/\psi\eta$ from the charmed meson loops will be smaller than that from the charmed-strange meson loops. The kinematic regions of the ATS are also different for charmed and charmed-strange meson loops.

\section{Summary}
In this work, to hunt for the charmonium-like states with hidden $s\bar{s}$, we investigate the radiative transition processes $e^+e^-$$\to$$\gamma J/\psi\phi$, $e^+e^-$$\to$$\gamma J/\psi\omega$ and the isospin violation process $e^+e^-$$\to$$\pi^0 J/\psi\eta$. These processes will receive contributions from the rescattering processes via the charmed-strange meson loops, of which the corresponding amplitudes are demonstrated to be very important. Especially, when the kinematics of these processes meets some special conditions, the ATSs can be present in the rescattering amplitudes, which will behave themselves as narrow peaks in the corresponding invariant mass distributions. This implies that the non-resonance explanation about the resonance-like structures is possible. The genuine particles, such as tetraquark states, molecular states and hybrids, may not be necessary to be introduced when describing the observations of some $XYZ$ particles. The ATS is just the kinematic singularity of the $S$-matrix elements, and
the locations of the resonance-like peaks induced by the ATS will mainly depend on the specific kinematic configurations. In our discussion, usually they will stay close to the $D_s^{+}D_s^{-}$, $D_s^{*+}D_s^{-}$ and $D_s^{*+}D_s^{*-}$ thresholds, which we call normal thresholds. Sometimes the discrepancy between the normal and anomalous thresholds can be larger. Taking into account the locations of the ATSs can move, this offers us a criterion to distinguish kinematic singularities from genuine resonances, because the peak positions of the genuine resonances are usually thought to be relatively stable. However, although the kinematic regions of the ATS for the charmed-strange meson loops are sizable, they are not too large. To observe the movement of the ATS, the higher energy resolution of the experiments is necessary.

\subsection*{Acknowledgments}
We thank the helpful discussions with Q. Zhao and G. Li.
This work was supported by the Japan Society for the Promotion of Science under Contract No. P14324, and the JSPS KAKENHI (Grant No. 25247036).

\appendix
\section{Quark-interchange model}

\begin{figure}[htb]
	\centering
	\includegraphics[width=0.7\hsize]{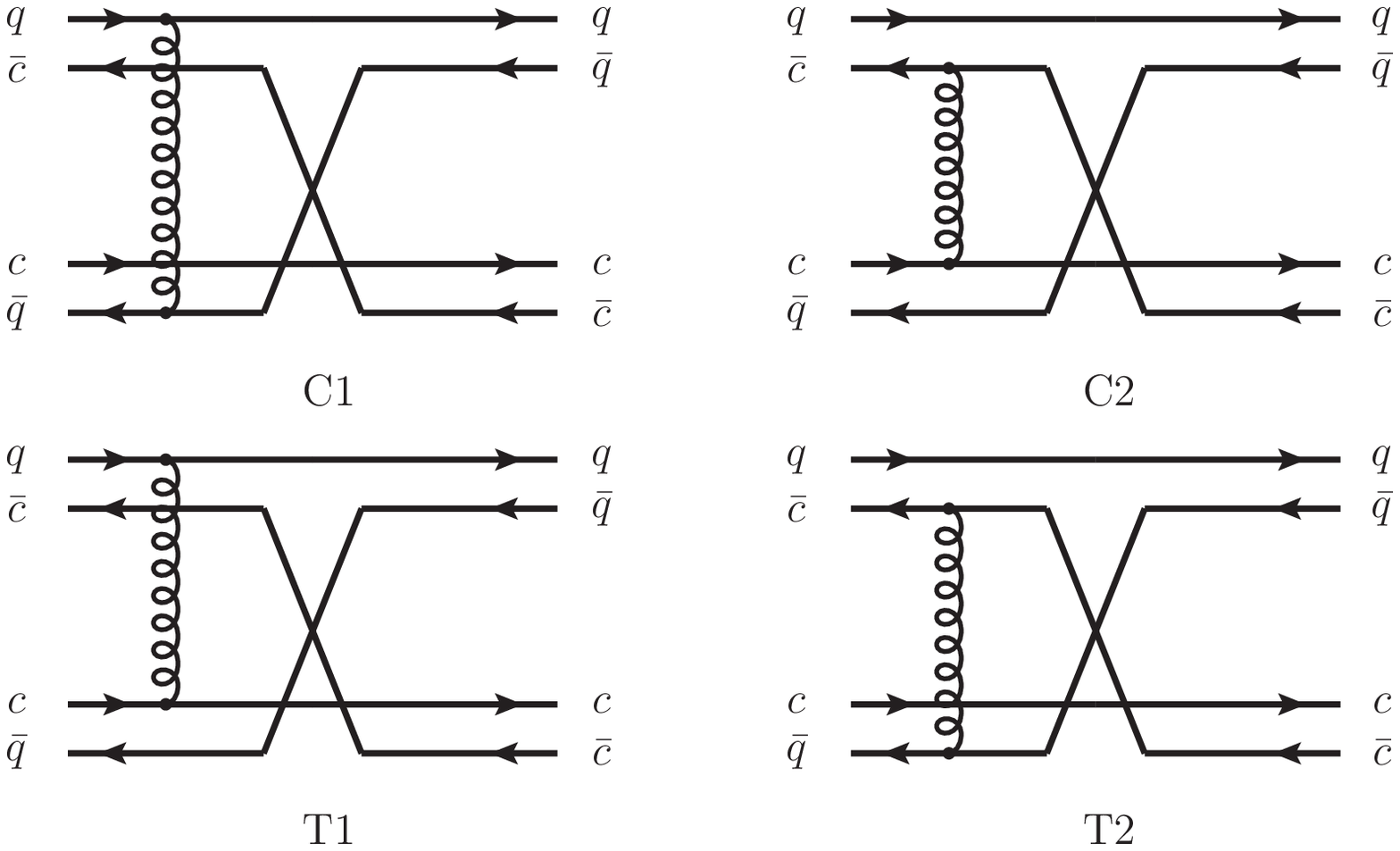}\\
	\caption{Quark-interchange diagrams contributing to the anticharmed meson-charmed meson scattering into the light flavor meson and the charmonia.}\label{priordiagram}
\end{figure}

In the reactions $D_s^{(*)+}D_s^{(*)-}$$\to$$J/\psi\phi$, $J/\psi\omega$ and $J/\psi\eta$, $c$ and $\bar{c}$ are recombined into
a charmonium state, which is governed by the short range
interaction. To describe these meson-meson scatterings at the quark
level, we will employ the Barnes-Swanson quark-interchange model to
estimate the transition
amplitudes~\cite{Barnes:1991em,Barnes:1992qa,Wong:1999zb,Barnes:2000hu,Wong:2001td,Barnes:2003dg,Liu:2014eka}.
In this approach, the non-relativistic quark potential model is
used, and the hadron-hadron scattering amplitudes are evaluated at
Born order with the interquark Hamiltonian. In the case of the
anticharmed meson-charmed meson scatterings, the amplitudes arise
from the sum of the four quark exchange diagrams as illustrated in
Fig. \ref{priordiagram}.
The interaction $H_{ij}$ between constituents $i$ and $j$ is
represented by the curly line in Fig.\ref{priordiagram}, and is
taken to be
\begin{eqnarray}
H_{ij} &\equiv& \frac{\mathbf{\lambda}(i)}{2}\cdot\frac{\mathbf{\lambda}(j)}{2} V_{ij}({r_{ij}})  = \frac{\mathbf{\lambda}(i)}{2}\cdot\frac{\mathbf{\lambda}(j)}{2} \left( V_{{conf}}+V_{{hyp}} +V_{constant} \right) \nonumber \\
&=& \frac{\mathbf{\lambda}(i)}{2}\cdot\frac{\mathbf{\lambda}(j)}{2} \bigg\{
\frac{\alpha_s}{r_{ij}}-\frac{3b}{4} r_{ij} -\frac{8\pi\alpha_s}{3m_i m_j}\mathbf{S}_i \cdot \mathbf{S}_j\  \left( \frac{\sigma^3}{\pi^{3/2}} \right) e^{-\sigma^2 r_{ij}^2} +V_{constant} \bigg\}. \label{hamiltonian}
\end{eqnarray}
This Hamiltonian contains a Coulomb-plus-linear confining potential $V_{conf}$ and a
short range spin-spin hyperfine term $V_{hyp}$, which is motivated by one gluon exchange.

The Born-order $T$-matrix element $T_{fi}$ can be expressed as the product of three
factors for each of the diagrams in Fig.\ref{priordiagram},
\begin{eqnarray}
T_{fi}=(2\pi)^3 I_{flavor}I_{color} I_{spin-space}.
\end{eqnarray}
Since there is no orbitally excited state involved in our
discussion, the factor $I_{spin-space}$ can be further factored into
\begin{equation}
I_{spin-space}=I_{spin} \times I_{space}.
\end{equation}
The space factors are evaluated as the overlap integrals of the
quark model wave functions. It is convenient to write these overlap
integrals in the momentum-space. For the four diagrams of
Fig.~\ref{priordiagram}, in the reaction $AB\to CD$, where $AB$ and
$CD$ are the initial and final meson pairs respectively, the space
factors read
\begin{eqnarray}
&& I_{space}^{C1} = \int\int  d\mathbf{k}\ d\mathbf{q}\  \Phi_A(2\mathbf{k})\ \Phi_B(2\mathbf{k}-2\mathbf{P_C})\ \Phi_C(2\mathbf{q}-\mathbf{P_C})\ \Phi_D(2\mathbf{k}-\mathbf{P_C}) \ V(\mathbf{k}-\mathbf{q}), \nonumber \\
&& I_{space}^{C2} = \int\int  d\mathbf{k}\ d\mathbf{q}\  \Phi_A(-2\mathbf{k})\ \Phi_B(-2\mathbf{k}-2\mathbf{P_C})\ \Phi_C(-2\mathbf{k}-\mathbf{P_C})\ \Phi_D(-2\mathbf{q}-\mathbf{P_C}) \ V(\mathbf{k}-\mathbf{q}),   \nonumber\\
&& I_{space}^{T1} = \int\int  d\mathbf{k}\ d\mathbf{q}\  \Phi_A(2\mathbf{k})\ \Phi_B(2\mathbf{q}-2\mathbf{P_C})\ \Phi_C(2\mathbf{q}-\mathbf{P_C})\ \Phi_D(2\mathbf{k}-\mathbf{P_C}) \ V(\mathbf{k}-\mathbf{q}),  \nonumber\\
&& I_{space}^{T2} = \int\int  d\mathbf{k}\ d\mathbf{q}\
\Phi_A(-2\mathbf{k})\ \Phi_B(-2\mathbf{q}-2\mathbf{P_C})\
\Phi_C(-2\mathbf{k}-\mathbf{P_C})\ \Phi_D(-2\mathbf{q}-\mathbf{P_C})
\ V(\mathbf{k}-\mathbf{q}), \nonumber \\ \label{eqprior}\end{eqnarray} 
where
$\bf{P_C}$ is the center-of mass momentum of meson $C$, and the
potential $V(\mathbf{p})$ is obtained via the Fourier transformation of
$V(r)$.

\end{document}